\documentclass[10pt,a4paper,referee]{aa}
\usepackage{aalongtable}
\usepackage[tight,scriptsize]{subfigure}
\usepackage{natbib}
\usepackage{graphicx}
\usepackage{amssymb}
\newcommand{\II}{\scriptsize{II}  \normalsize}
\begin{document}
\title{Long-term chromospheric activity of non-eclipsing RS CVn-type
    stars} \author{Andrea P. Buccino \and Pablo J. D. Mauas}
    \institute{ Instituto de Astronom\'\i a y F\'\i sica del Espacio
    (CONICET), C.C. 67 Sucursal 28, 1428-Buenos Aires Argentina}
    \offprints{A. P. Buccino, \email{abuccino@iafe.uba.ar}}
    \date{Received/ Accepted} \abstract{The IUE database provides a
    large number of UV high and low-resolution spectra of RS CVn-type
    stars from 1978 to 1996. In particular, many of these stars were
    monitored continuously during several seasons by IUE. }{ Our main
    purpose is to study the short and long-term chromospheric activity
    of the RS CVn systems most observed by IUE: HD 22468 (V711 Tau, HR
    1099, K1IV+G5V), HD 21242 (UX Ari, K0IV+G5V) and HD 224085 (II
    Peg, K2IV).}{We first obtain the Mount Wilson index $S$ from the
    IUE high and low-resolution spectra. Secondly, we analyse with the
    Lomb-Scargle periodogram the mean annual index $\langle S\rangle$
    and the amplitude of the rotational modulation of the index
    $S$.}{For HD 22468 (V711 Tau, HR 1099), we found a possible
    chromospheric cycle with a period of $\sim$18 years and a shorter cycle
    with a period of $\sim$3 years, which could be associated to a
    chromospheric ``flip-flop'' cycle. The data of HD 224085 (II Peg)
    also suggest a chromospheric cycle of $\sim$21 years and a
    flip-flop cycle of $\sim$9 years.  Finally, we obtained a possible
    chromospheric cycle of $\sim$7 years for HD 21242 (UX
    Ari).}{}
\titlerunning{Long-term chromospheric activity of RS CVn-type stars}
\authorrunning{Buccino \& Mauas}
\maketitle
\section{Introduction}\label{sec.intro}

One of the principal diagnostics for solar
and stellar chromospheric activity is the emission in the Ca
\scriptsize{II} \normalsize H and K resonance lines (at 3968 and 3934
\AA). In particular, the largest dataset of activity measurements
available at present, comprising observations of over 2000
stars, is the one obtained with the Mount Wilson HK
spectrophotometers, which have been used since 1966 to measure high-precision Ca \scriptsize{II}
\normalsize H+K fluxes. As an indicator of
stellar activity, an index $S$ has been defined as the ratio between the
line-core fluxes and the flux in the continuum nearby.

Since the Mg \II  h and k lines (at 2803 and
2796 \AA\-)  are
formed in a similar way to the Ca \scriptsize{II} \normalsize lines,
they are  also good
indicators of the thermal structure of stellar atmospheres, specially
from the high photosphere to the upper part of the chromospheric
plateau.

The \emph{International Ultraviolet Explorer (IUE)} provides a large
 database of low and high-resolution UV spectra in the band 1150-3350
 \AA\-, from 1978 to 1996, which can be used to study long-term
 stellar activity.

In a previous work (\citealt{2008A&A...483..903B}, hereafter Paper I)
    we obtained a colour dependent relation between the Mount Wilson
    index $S$ and Mg \II line-core surface fluxes derived from IUE
    high-resolution spectra for F5 to K3 main sequence stars.  As an
    application of this calibration, we computed the Mount Wilson
    index for all the dF to dK stars which have
     high-resolution IUE spectra. In that paper we also
    combined IUE observations with those obtained by
    \cite{1996AJ....111..439H}, and by our group (see
    \citealt{2004A&A...414..699C} and
    \citealt{2007astro.ph..3511C}). For some of the most frequently
    observed main sequence stars, we combined the Mount Wilson index
    $S$ from the IUE spectra, together with the ones derived from
    visible spectra, covering the period between 1978 and 2005.
      Analysing the data with the Lomb-Scargle periodogram, we
    were able to confirm the chromospheric activity cycle of
    $\epsilon$ Eri (HD 22049) and $\beta$ Hydri (HD 2151) with periods
    of $\sim$5 and $\sim$12 years respectively, and we found a
    magnetic cycle in $\alpha$ Cen B (HD 128621) of $\sim$8 years.

Similarly,  the purpose of the present study is to calibrate
    IUE low-resolution observations of Mg \II h+k lines with the Mount
    Wilson index $S$, as a complement of the work presented in Paper
    I, and to apply this calibration to study the activity in RS CVn stars.

RS CVn-type stars are well known due to their strong
chromospheric plages, coronal X-ray, and microwave emissions, as well
as strong flares in the optical, UV, radio, and X-ray.  These stars
belong to a class of close detached binaries where the more massive
primary component is a G-K giant or subgiant and the secondary is a
subgiant or dwarf of spectral classes G to M.  The magnetic activity
signatures are generally dominated by the primary star of the
system. Since these stars are fast rotators enforced by tidal
synchronization, they are, in general, more active than single stars
of similar mass and age.

The remarkable activity and high luminosity of these stars make them
favourite targets for light curve modeling
\citep{1986A&A...165..135R,1995A&A...293..107D,1999A&A...350..626B,2007ApJ...659L.157B},
Doppler imaging
\citep{1992A&A...265..682D,1997MNRAS.291..658D,1999MNRAS.302..457B,2000MNRAS.318.1171K}
and spectral line analysis
\citep{1987A&A...186..241W,1996ApJ...470.1172D,2000A&AS..146..103M,2001A&A...365..128P}. In
general, most long-term stellar activity studies of RS CVn-type stars
are based on easily detected optical photometric variations
produced by their long-lived large spots.  Recently,
\cite{2008A&A...480..495M} presented the results of a long-term
photometric monitoring project, where he carefully  studied the
magnetic activity of 14 late-type active components of close binary systems
and its evolution on different time scales.
\cite{1998A&A...332..541L,2006A&A...455..595L} and
\cite{2007ApJ...659L.157B} presented similar long-term studies for
individual RS CVn stars.

In general, these studies concluded that the mean activity level of RS
CVn-type stars presents cyclic variations similar to the 11-year solar
cycle. Moreover, they found that large starspots tend to
appear always at particular longitudes, called ``active longitudes''
(ALs), which are generally separated by 180$^\circ$ on average
and differ in their level of activity.  \cite{1998A&A...336L..25B}
found that the dominant activity switches periodically from one active
longitude to the other in most RS CVn stars. This type of stellar
activity cycle, known as ``flip-flop cycle'', is mainly related to the
non-axisymmetric redistribution of the spotted area.


The IUE database provides a large number of UV high and low-resolution
spectra of RS CVn-type stars. Furthermore, IUE monitored these stars
continuously during several seasons. In most cases, the observations
were aimed at uniform coverage with respect to orbital phase over more
than one rotation period in each season
(e.g. \citealt{1996ApJ...470.1172D},
\citealt{1999A&A...350..571B}). In this work we analyse the short and
long-term chromospheric activity of the non-eclipsing RS CVn stars
most observed by IUE and we compare the chromospheric and photospheric
patterns of variability.

In particular, in Section \S\ref{sec.calibbaja}, we derive a relation
 to determine the Mount Wilson index from the Mg \II line-core fluxes
 measured on IUE low-resolution spectra. As an application of this
 calibration, in Section \S\ref{sec.rscvn} we study the Mount Wilson
 indices we derived from both IUE high and low-resolution
 spectra of the RS CVn stars HD 22468 (V711 Tau, HR 1099),  HD 21242
 (UX Ari) and HD 224085 (II Peg).

\section{Mg \II lines and Mount Wilson index}\label{sec.calibbaja}

\begin{figure}[htb!]
\centering
\includegraphics[width=0.77\textwidth]{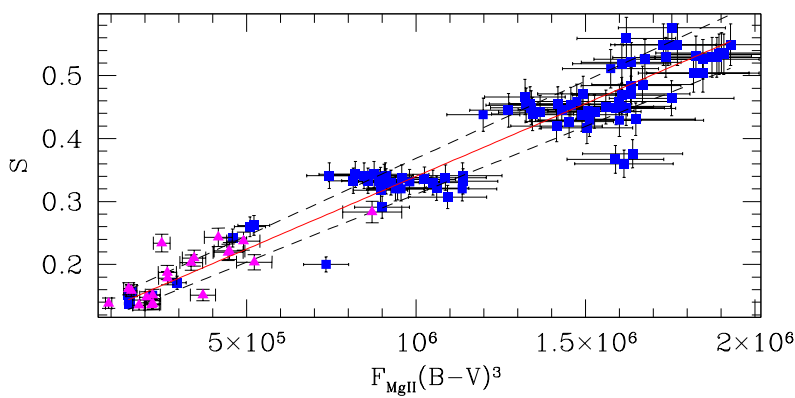}
\caption{$S$ vs. $F_{Mg II}(B-V)^{3}$ for the set of 117
    nearly simultaneous observations of Mount Wilson index and IUE
    high-resolution Mg \II \footnotesize h and k fluxes for dwarfs we
    used in Paper I ($\blacksquare$). We considered errors of 10\% for
    $F_{Mg\II}$ and 6\% for $S$. The least square fit (solid line) has
    a correlation coefficient of 0.95.  The dotted lines indicate
    $\pm$3$\sigma$ from the fit. The triangles ($\blacktriangle$)
    represent non-simultaneous Mount Wilson indices and IUE
    high-resolution Mg \II \footnotesize fluxes of cool sub-giant
    stars, not used in the calibration.}\label{fig.calibS_alta}
\end{figure}

  In Paper I we inter-calibrated the index $S$ and Mg \II
  fluxes using 
  quasi-simultaneous IUE high-resolution observations  
  for a set of dwarf stars with spectral types F to K. Since in the
  present work we intend to use this calibration to study the activity
  of  RS CVn systems, which have sub-giants as their primary star,
  we check the accuracy of applying this calibration to cool
  sub-giants. To do so, we obtained the IUE Mg \II fluxes for those
  cool sub-giant stars also observed at Mount Wilson Observatory. In
  Fig. \ref{fig.calibS_alta} we show the calibration of Paper I, and
  we include these sub-giants. It can be seen that these stars follow
  this calibration within the statistical errors.

Since the IUE database also provides a large
  number of low-resolution spectra of late-type stars, in
  particular RS CVn stars, it is important to incorporate to the
  systematic studies of magnetic activity. To this end, in what
  follows we analyse the relation between the Mg \II line-core
  fluxes derived from IUE low-resolution spectra and the Mount Wilson
  index.

\subsection{Observations}
\begin{figure}[htb!]
\centering
\includegraphics[width=0.7\textwidth]{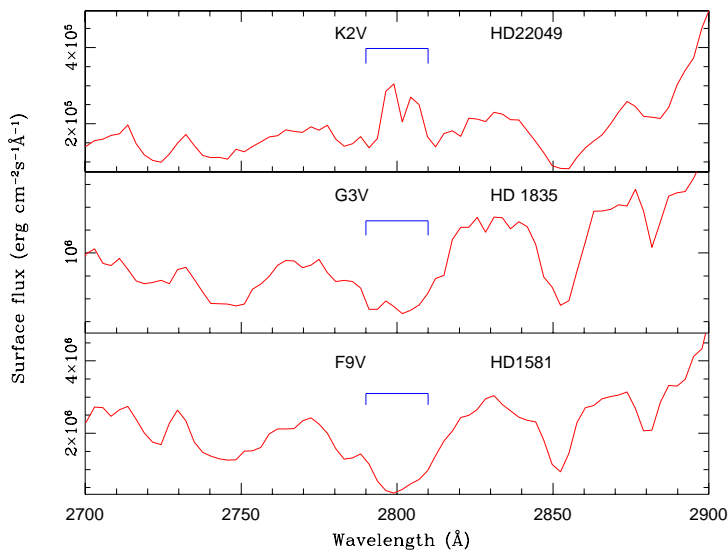}
\caption{IUE low-resolution  spectra of three representative stars: HD
  22049 (K2V), HD 1835 (G3V) and HD 1581 (F9V). Surface flux
  is obtained according to Eq. \ref{ec.fluxcal}. The   windows used to integrate the
  line-core  emission (2790 - 2810 \AA)\- are marked on each
  spectrum.}\label{fig.esp_b}
\end{figure}

In Fig. \ref{fig.esp_b} we present some examples of IUE low-resolution
 spectra of three F, G and K main sequence stars. These spectra
 present a resolution of $R=400$ at 2700 \AA, they are available from
 the IUE public library (at
 \textsf{http://sdc.laeff.esa.es/cgi-ines/IUEdbsMY}), and have been
 calibrated using the NEWSIPS (New Spectral Image Processing System)
 algorithm \citep{1997IUENN..57....1G}.  The internal accuracy of the
 low-resolution flux calibration is 10-15\%
 \citep{2000ApJS..126..517M}.

 The
observed  flux $f$ in the IUE spectra  was transformed to the surface
flux  \emph{F} using \citeauthor{1992A&A...258..432S}'s
(\citeyear{1992A&A...258..432S}) relation: 
\begin{equation}
log\,(F/f)=0.35+0.4(m_V+BC)+4\,\,log\,T_{eff},\label{ec.fluxcal}
\end{equation}
 where $m_V$ is the visual apparent magnitude  (from
the \emph{Hipparcos and Thycho Catalogue}
\citealt{1997A&A...323L..49P}, \citealt{1997A&A...323L..57H}), $BC$ is the
 bolometric  correction (from \citealt{1996ApJ...469..355F}) and
 $T_{eff}$  is
 the  effective temperature (from \citealt{1981ARA&A..19..295B}).

In Fig. \ref{fig.esp_b} we also show the windows used to integrate the
Mg \II line-core fluxes, which are 10 \AA-wide and are centred at
2795 \AA. The position and the width of these windows were chosen to
guarantee that the contribution of the integrated flux is
predominantly chromospheric, beyond the basal contribution. However,
it can be seen in the figure that the Mg \II line-core emission can only be
detected in stars of late spectral type, in which the level of
phostospheric continuum is low. Therefore,  the Mg \II
line-core fluxes derived from IUE low-resolution spectra could be only
considered as an activity indicator in  stars cooler than G3.

On the other hand, the ``IUE Newly Extracted Spectra'' (INES) system
  provides a \emph{rebinned} spectrum for each high-resolution image,
  which is obtained by re-sampling the high-resolution spectrum into
  the low-resolution wavelength domain. The re-sampling has been
  performed so that the total flux is conserved. However, the rebinned
  spectra have not been convolved with the low-resolution Point Spread
  Function (PSF) and, in consequence, have a better spectral
  resolution than low-dispersion spectra \citep{2000A&AS..141..343G}.

\begin{figure}[htb!]
\centering
\includegraphics[width=0.7\textwidth]{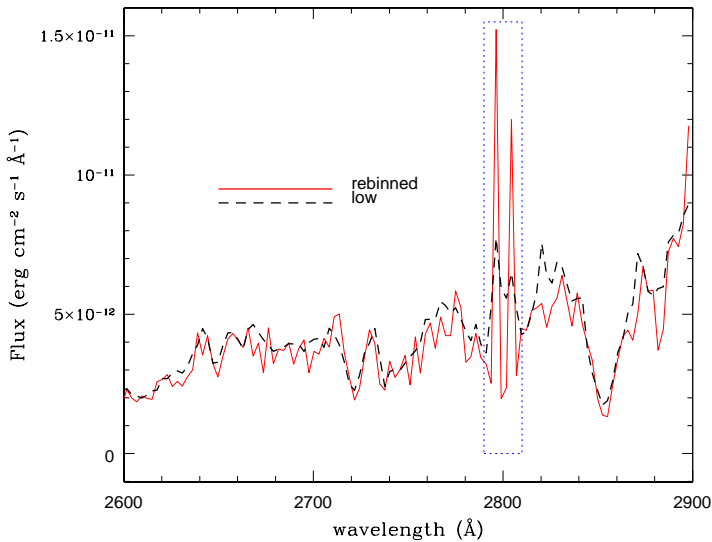}
\caption{IUE low-dispersion (dashed line) and
  rebinned (solid line) spectra of HD 22049, both  observed on January,
  14$^{\textrm{th}}$  1984 with a time difference of  1 hour.  }\label{fig.comp_a_b}
\end{figure}

In Fig. \ref{fig.comp_a_b} we plot two quasi-simultaneous rebinned and
 low-dispersion spectra of the star HD 22049 (K2V). We can see that both
 spectra are  similar but not equal. However, the Mg \II
 line-core fluxes integrated between 2790 and 2810 \AA\- (dotted rectangle) on both spectra only
 differ by 1-2\%.  Therefore, in order to calculate the Mount Wilson
 index from the IUE spectra, we can consider the low-resolution and the rebinned
 spectra indistinguishable.

\subsection{Calibration of $S$ vs. $F^r_{Mg II}$}

To intercalibrate the Mount Wilson index $S$ and the Mg \II line-core
surface fluxes derived from the IUE low-resolution spectra, we used
the same set of quasi-simultaneous observations used in Paper I (see
Fig. \ref{fig.calibS_alta}). In particular, we measured the Mg \II
fluxes on the IUE \emph{rebinned} spectra ($F^r_{Mg II}$)
corresponding to the high-resolution spectra used in our previous
work.

\begin{table}[htb!]
\caption{Stars used in the Mg \scriptsize{II} \normalsize -
and the index \emph{S} calibration. }\label{tab.estcal}
\centering
\begin{tabular}{r c r r r}
\hline
Stars &  Spectral &       &      & $T_{eff}$\\
 HD &   Class and &  $m_V$& $B-V$    & (K)\\
    &   Type      &       &        &    \\
\hline
\hline
  1835 & G3V & 6.39 & 0.66 & 5675\\
 10700 & G8V & 3.50 & 0.72 & 5500\\
 17925 & K0V & 6.04 & 0.87 & 5170\\
 20630 & G5V & 4.83 & 0.68 & 5610\\
 22049 & K2V & 3.73 & 0.88 & 5140\\
 26965 & K1V & 4.43 & 0.82 & 5295\\
101501 & G8V & 5.33& 0.72 & 5500\\
115404 & K3V & 6.52 & 0.94 & 4990\\
131156 & G8V & 4.72 & 0.72 & 5500\\
149661 & K0V & 5.75 & 0.82 & 5295\\
152391 & G8V & 6.64 & 0.76 & 5295\\
\hline
\hline
\end{tabular}

\end{table}

As mentioned before, the Mg \II line-core fluxes can be considered as
an activity indicator only in late-type stars, where the line-core
emission is remarkably larger than the contribution of the continuum
nearby. Therefore, we reduced our set of calibration stars to those
with \mbox{$B-V>0.65$}, listed in Table \ref{tab.estcal}.

We proposed a calibration for $S$ vs. $F^r_{MgII}$ analogous to the
one obtained in Paper I for the Mg \II fluxes derived from IUE high-resolution spectra:
\vspace{-0.1cm}
\begin{equation}
S = a\,(B-V)^\alpha\,F^r_{Mg II}+b\,,\label{eq.calibS_baja}
\end{equation}
with  $F^r_{Mg\II}$ expressed in
erg cm$^{-2}$ s$^{-1}$. We found that the best correlation coefficient R=0.82 is obtained for
$\alpha$=(5.65 $\pm$ 0.20). 

\begin{figure}[htb!]
\centering
\includegraphics[width=0.8\textwidth]{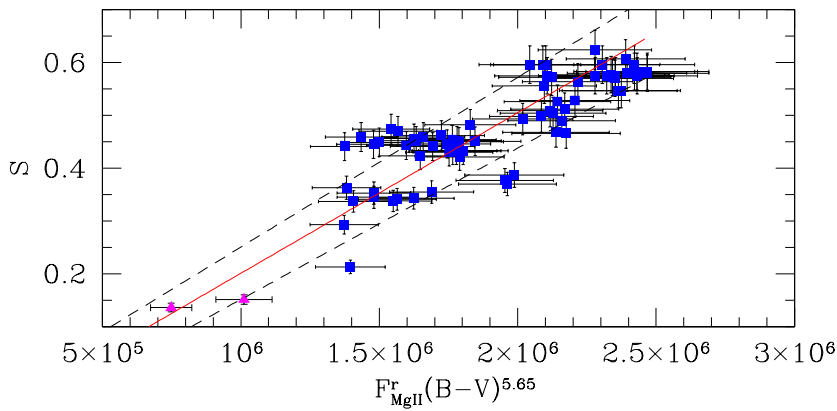}
\caption{$S$ vs. $F^r_{Mg II}(B-V)^{5.65}$ for the stars listed in
  Table \ref{tab.estcal}. We considered errors of 10\% for
  $F^r_{Mg\II}$ and 6\% for $S$. The least square fit (solid line) has
  a correlation coefficient of 0.82.  The dotted lines indicate the
  points that apart $\pm\sigma$ from the fit. The triangles
  ($\blacktriangle$) represent non-simultaneous Mount Wilson indices
  and IUE high-resolution Mg \II \footnotesize fluxes of cool
  sub-giant stars with $B-V>0.65$, not used in the calibration.
  }\label{fig.calibS_baja}
\end{figure}

 The uncertainties in the determination of the effective temperature
  and the bolometric correction of the star may introduce a $\sim$10\%
  error in the Mg \II surface fluxes obtained with
  Eq. \ref{ec.fluxcal} \citep{1992A&A...258..432S}.  On the other
  hand, the dispersion in $F^r_{Mg\II}$, due to the fact that there is
  a single value of $S$ for IUE spectra differing in less than 36
  hours, has a standard deviation that can be up to 5\%.  The standard
  deviations of the Mount Wilson indices $S$ obtained from
  \cite{1991ApJS...76..383D} do not exceed 10\% of its mean
  value. Therefore, we varied the error of $F^r _{Mg\II}$ and $S$ near
  the maximum standard deviation (15\% for $F^r _{Mg\II}$ and 10\% for
  $S$) in order to minimize the $\chi^2$-statistic.

Considering  errors in both coordinates  and
minimizing the expression:
\begin{equation}
\centering
\chi^2=\Sigma_{i=1}^N\frac{(y_i-b-ax_i)^2}{\sigma_{y_i}^2+a^2\sigma_{x_i}^2},
\label{ec.chi_xy}
\end{equation}
where $y\equiv S$ and $x\equiv F^r_{Mg\II}(B-V)^\alpha$, we obtained
that the best parameters are \mbox{$a=(2.76\pm0.18)\times 10^{-7}$}
and \mbox{$b=-0.155\pm 0.041$}, which give a reduced
$\chi^2=1.27$ for uncertainties of 10\% and 6\% in the $F^r _{Mg\II}$
and $S$ values respectively.

In Fig. \ref{fig.calibS_baja} we plot the Mount Wilson index $S$
vs. the Mg \scriptsize{II} \normalsize line-core surface fluxes for
dwarf stars and the best linear fit given by
Eq. \ref{eq.calibS_baja}. As in Fig. \ref{fig.calibS_alta}, we
also included the index $S$ and the Mg \II fluxes for cool
sub-giant stars with $B-V>0.65$. We can see that these sub-giants
follow the calibration obtained in Eq. \ref{eq.calibS_baja} within the
statistical errors.

\section{RS CVn  stars}\label{sec.rscvn}

The IUE database provides a large number of UV high and low-resolution
spectra of RS CVn-type stars. In this section, we analyse the long and
short-term chromospheric activity of three RS CVn stars observed many
times by IUE. In particular, we analysed the activity signatures,
which mainly correspond to the primary star, of the system: HD 22468
(V711 Tau, HR 1099), HD 21242 (UX Ari, HD 21242) and HD 224085 (II
Peg).

 Most observations were performed at high-resolution. To add to
  these observations the low-resolution ones, we transform the Mg \II
  fluxes to $S$, following the calibrations obtained here and in Paper
  I. We note that the values obtained from low-resolution observations
  have larger uncertainties since they include an extra activity
  component related to the continuum nearby, whose variation could be
  up to 30\% for very active stars (see \S2 in Paper I). We also note
  that, for a given star, the error in the transformation to surface
  fluxes (Eq. \ref{ec.fluxcal}) is common to all the observations and,
  therefore, does not affect our results regarding stellar
  variability, although it might affect the mean level of activity.

In this way, our observations cover the period between 1978 and 1996.
Even if, in several cases, the density of measurements along the years
is low, we can infer the average level of activity and its
variability for the whole period and during a few short
intervals of time.

For the stars mentioned above, we
list in Table \ref{tab.timescalevar} the average, maximum and
minimum level of activity reached along decades (Cols. 4 and 5) and the
variations recorded in particular years (Col. 6).

\begin{table*}[p!]
\centering
\begin{minipage}{\textwidth}
\caption{Long and short term variability records of the RS
 CVn-type stars most observed
 by IUE.}\label{tab.timescalevar}
\begin{tabular}{c c  l r r c c  c l c l }
\hline
Star  & HD &Sp. type &$m_V$ & $B-V$&$P_{rot}$\footnote{References for the stellar  rotation period $P_{rot}$:
     (i) \cite{1983ApJ...268..274F},
  (ii) \cite{2001A&A...370..974D}.} & $i$\footnote{References for the inclination angle of the system:
     (1) \cite{1993A&AS..100..173S}, (2) \cite{1998A&A...334..863B}.}  & Mean\footnote{Mean
  annual Mount Wilson index for the period 1978-1996.} &  Long  scale \footnote{Maximum and minimum level of activity reached along
decades.}  & Short scale \footnote{Variations recorded in particular years.} &  Year \\
 Name  & &\& class & & & (days)&(deg) &activity& variability & variability&     \\
&  & & & & & &$\langle S\rangle$  &($\sim$ decades)& ($\sim$ months) &\\
 &  & & & &  & & &$S_{min}$/$S_{max}$&  $S_{min}$/$S_{max}$ &\\
\hline
\hline
V711 Tau  & 22468 &K1IV+G5V&5.82 &0.85& 2.837$^\textrm{\scriptsize{(i)}}$&33$^\textrm{\scriptsize{(1)}}$ &1.864 &1.651/2.090&1.700/2.396&1981\\
        & &   & &  &  & &  &   & 1.798/2.237 &1982\\
        & &   & &  &  & &  &   & 1.771/2.617 &1984\\
        & &   & &  &  & &  &   & 1.286/3.295&1986\\
        & &   & &  &  & &  &   &  1.431/2.814&1989\\
        & &   & &  &  & &  &   & 1.702/2.173 &1990\\
        & &   & &  &  & &  &   &  1.558/1.851&1991\\
        & &   & &  &  & &  &   & 0.794/2.785 &1992\\
        & &   & &  &  & &  &   &1.668/2.908  &1993\\
        & &   & &  &  & &  &   &1.406/2.751 &1994\\
UX Ari & 21242 & K0IV+G5V & 6.47&0.90 &6.483$^\textrm{\scriptsize{(ii)}}$&59$^\textrm{\scriptsize{(1)}}$ &1.277&1.115/1.480&1.132/1.501
     &1978    \\
     & & &  &  & & &  & &   0.994/5.056  &1979    \\
     & & &  &  & & &  & &   1.284/1.709  &1985    \\
     & & &  &  & & &  & &    1.026/1.626 &1987    \\
     & & &  &  & & &  & &   1.018/1.386 &1988    \\
     & & &  &  & & &  & &   1.061/1.645  &1990    \\
     & & &  &  & & &  & &   1.010/1.666 &1991    \\
     & & &  &  & & &  & &   1.140/1.753  &1994    \\
     & & &  &  & & &  & &   1.033/1.211  &1996    \\
II Peg & 224085 & K2IV &7.51 &1.01& 6.724$^\textrm{\scriptsize{(ii)}}$ & 60$^\textrm{\scriptsize{(2)}}$ & 1.694 &1.033/1.938&0.767/3.205& 1979\\
     & & &  &  & & &  &  &  0.477/3.775  &1980    \\
     & & &  &  & & &  &  &  1.323/2.390  &1981    \\
     & & &  &  & & &  &  &  1.227/3.545  &1983    \\
     & & &  &  & & &  &  &  1.627/1.821  &1984    \\
     & & &  &  & & &  &  &  1.817/2.005  &1985    \\
     & & &  &  & & &  &  &  1.379/1.973  &1986    \\
     & & &  &  & & &  &  &  1.641/2.483  &1989    \\
     & & &  &  & & &  &  &  1.389/2.434  &1990   \\
     & & &  &  & & &  &  &  1.497/3.280  &1992   \\
     & & &  &  & & &  &  &  1.600/2.226  &1993   \\
\hline
\hline
\end{tabular}
\end{minipage}
\end{table*}

In what follows we plot the index $S$ vs. time and we analyse the 
 stellar magnetic behaviour of each system in detail.  

\subsection{HD 22468 - V711 Tau - HR 1099}\label{subsec.hd22468}
HD 22468  is one of the most active RS CVn non-eclipsing
spectroscopic binary systems, consisting of a K1 subgiant primary and a G5 dwarf secondary in a
2.837-day orbit.

\begin{figure}[ht!]
\centering
\includegraphics[width=0.6\textwidth]{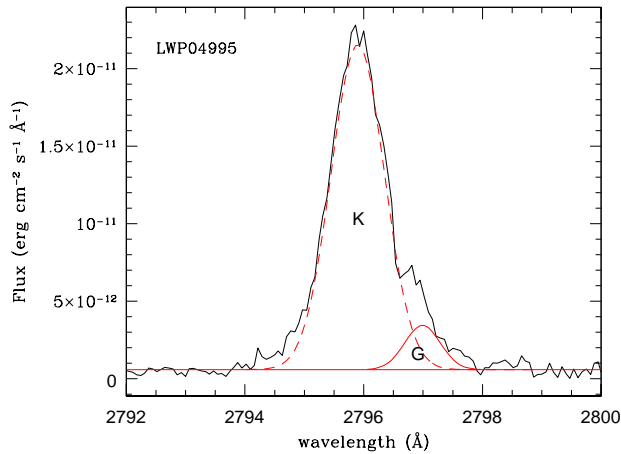}
\caption{HD 22468-V711 Tau-HR 1099. Mg \II k line fitted with two Gaussian
  functions, the dashed and solid curves represent the contribution of the
  primary and  secondary star of the system respectively.}\label{fig.gaussfit}
\end{figure}

In RS CVn stars the activity contribution of the secondary
  star is another source of error for determining the Mg \II line-core
  emission of the primary star. To estimate this contribution, we
  fitted the Mg \II h and k lines with two Gaussian functions to
  approximate the line-core emission of the primary and the secondary
  star in each line, using the IRAF task \emph{splot} (see
  Fig. \ref{fig.gaussfit}).

For all the IUE high-resolution spectra of HD 22468, we integrated the fluxes
under these curves and we found that the Mg \II
h+k line-core fluxes emitted by the G star do not exceed 10\% of the
total flux, which is within the standard
deviation we assumed for the measured flux. Therefore, the Mg \II
h+k line-core fluxes derived from the IUE high and low-resolution
spectra of HD 22468 can be considered as an activity indicator of the
K star of the system.

\begin{figure}[htb!]
\centering
\includegraphics[width=0.9\textwidth]{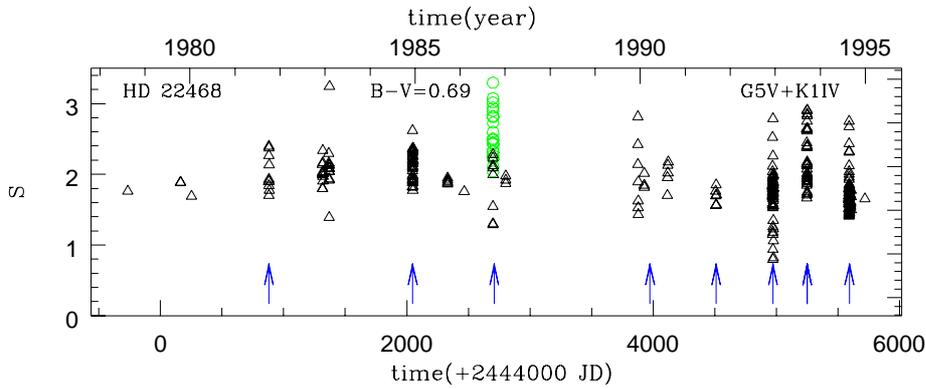}
\caption{HD 22468 - V711 Tau - HR 1099.  We indicate the Mount Wilson index $S$
obtained from the IUE high  and low-resolution  spectra with triangles
($\triangle$) and  circles ($\bigcirc$) respectively. We
  indicate with arrows the sets of data for which we analysed the short-term
  variability.}\label{fig.hd22468}
\end{figure}

In Fig. \ref{fig.hd22468} we plot the indices $S$  for this star, for
which we obtained a mean Mount Wilson index $\langle S \rangle$=1.864
from 1978 to 1995.  

\begin{figure}[htb!]
\centering
\subfigure[\label{hd22468_flare_1}]{\includegraphics[width=0.33\textwidth]{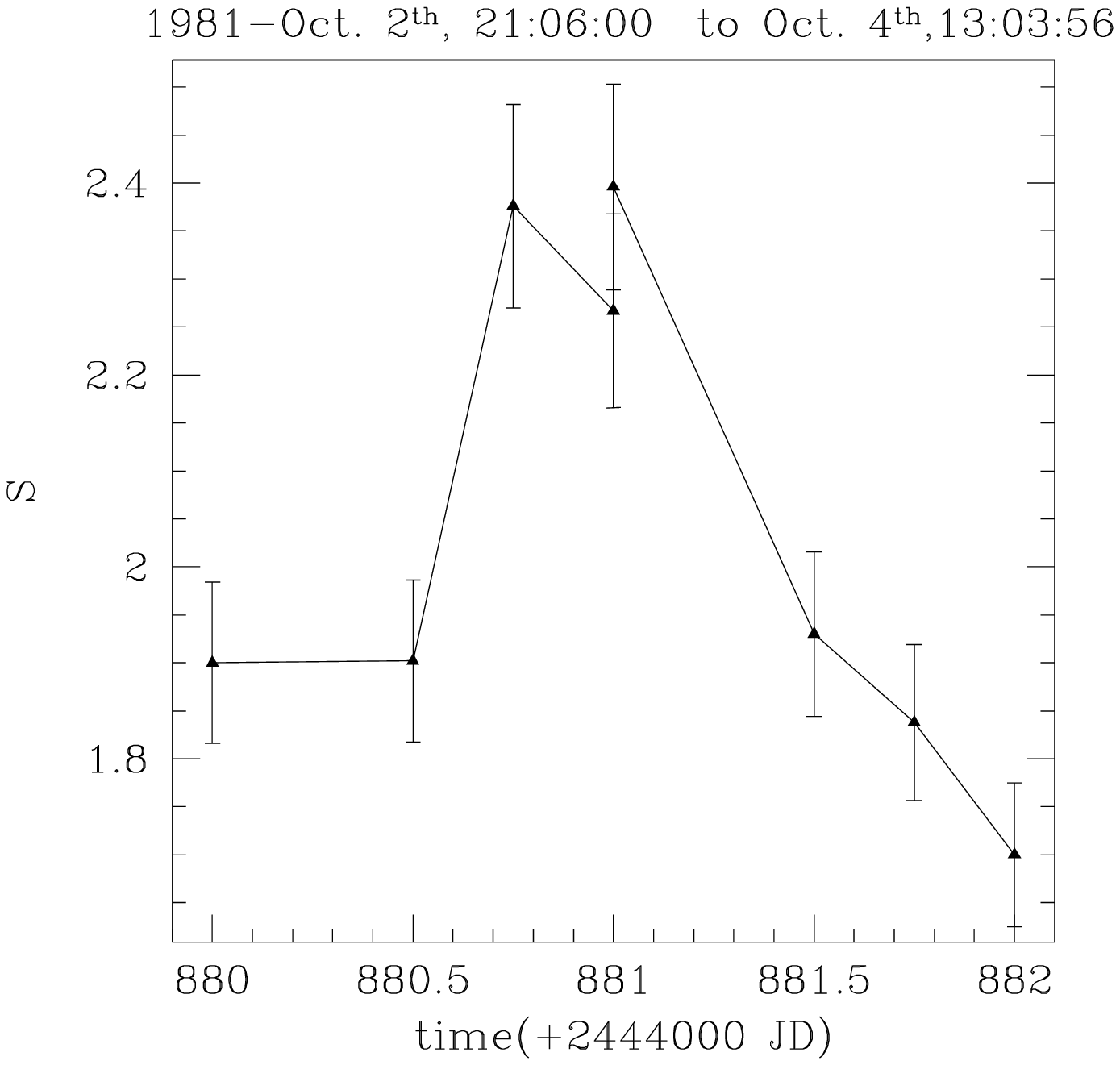}}\hfill
\subfigure[\label{hd22468_flare_2}]{\includegraphics[width=0.33\textwidth]{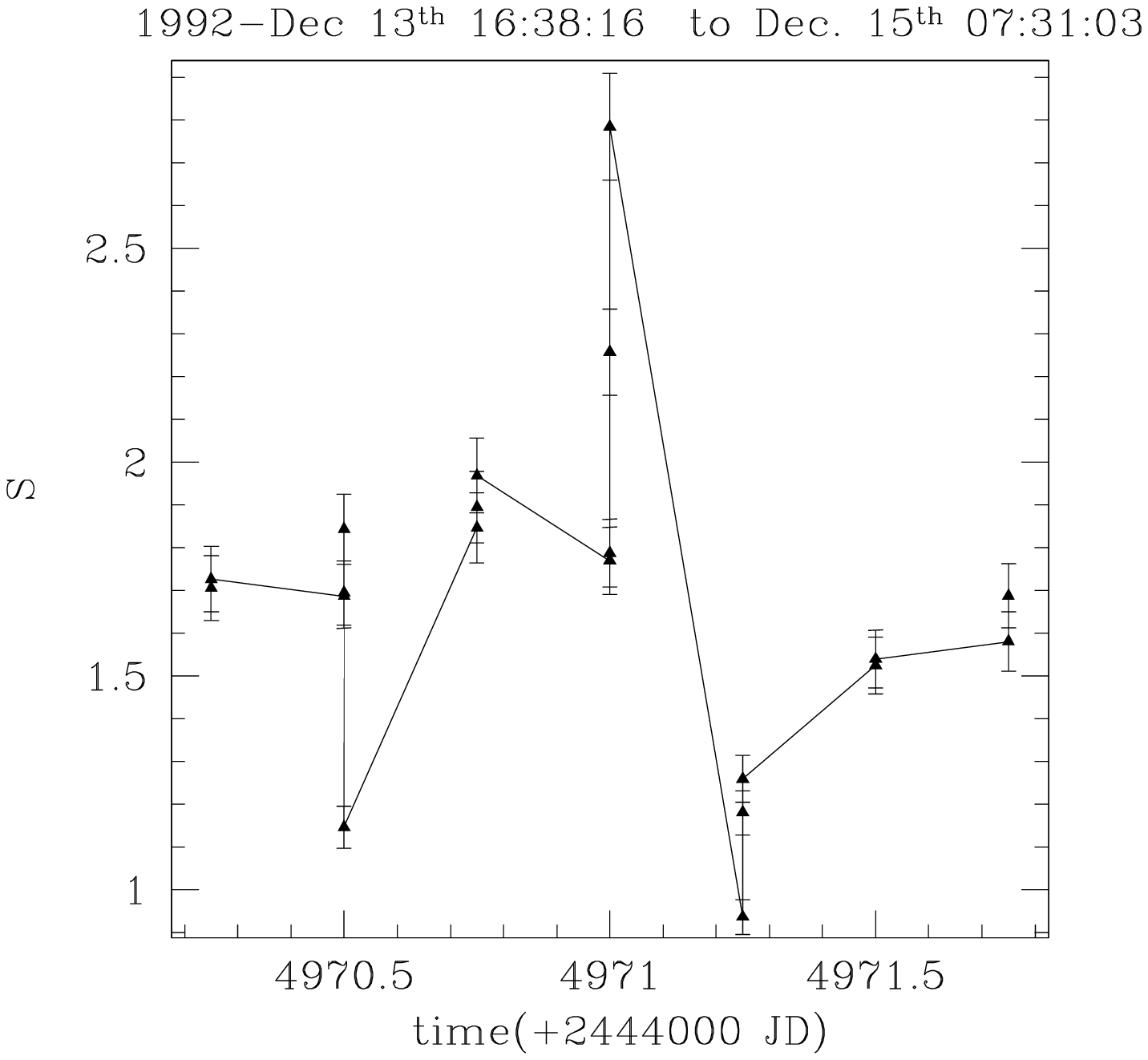}}
\subfigure[\label{hd22468_flare_3}]{\includegraphics[width=0.33\textwidth]{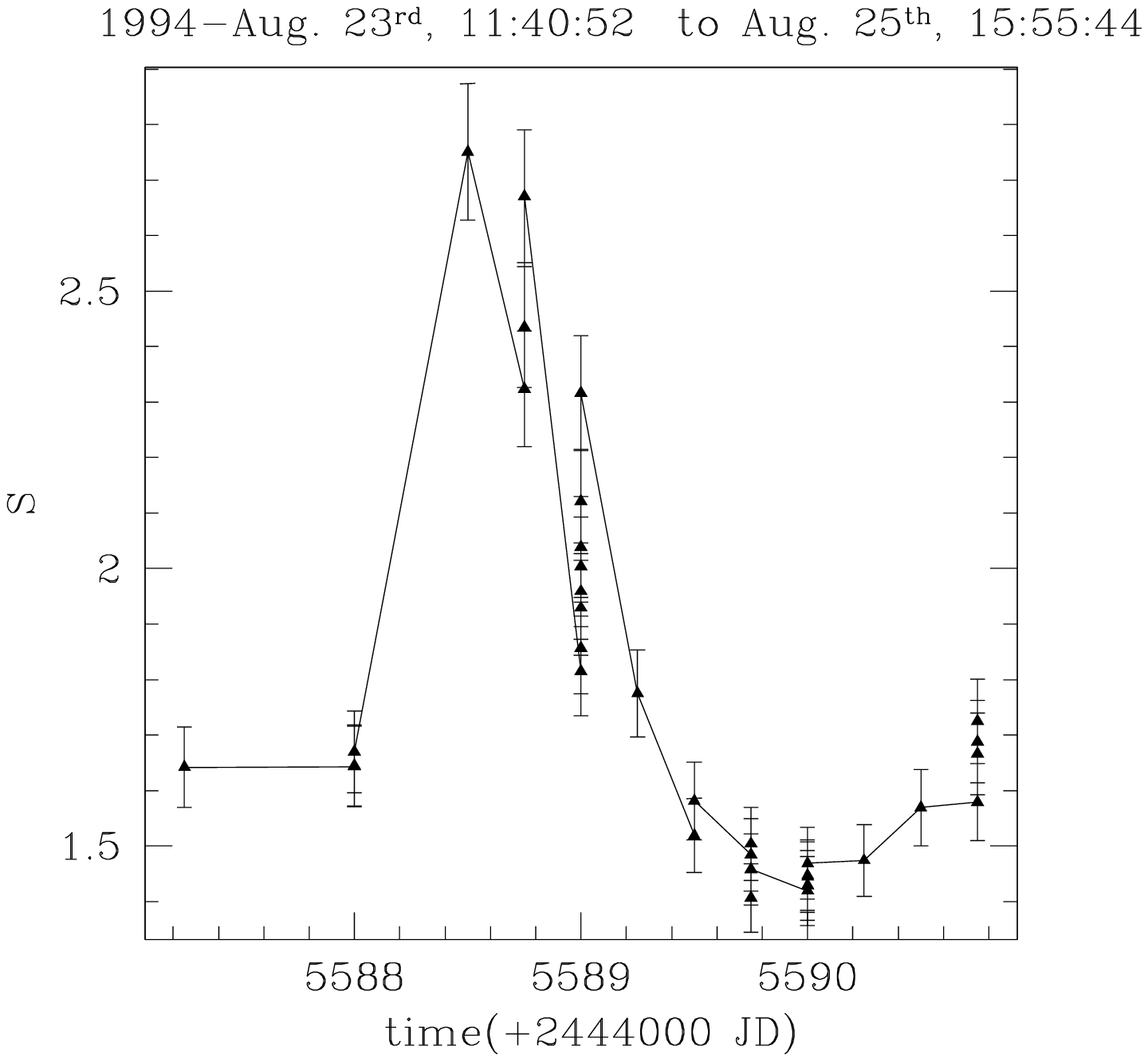}}\hfill
\caption{HD 22468 - V711 Tau - HR 1099. Three flares
    on 1981, Oct. 3$^{\textrm{rd}}$; 1992, Dec. 14$^{\textrm{th}}$ and
    1994, Aug. 24$^{\textrm{th}}$.}\label{fig.hd22468_flare}

\end{figure}

\cite{1992A&A...256..185C} reported that HD 22468
had fluxes comparable to those of the brightest solar regions and even
approaching those observed in solar flares. In fact, several flares
were observed with IUE, and can be noticed in Fig. \ref{fig.hd22468}. In
particular,  \cite{1989A&A...211..173L} reported a flare  on 1981,
October 3$^{\textrm{rd}}$, which released a total
energy of $\sim$ 10$^{32}$ erg. Using our calibration, we obtain that
$S$ increased 21\% in a few hours during that event.
\cite{1994AAS...185.8519B} reported another flare on 1994, August
24$^{\textrm{th}}$. During that date, the Mount Wilson index presented
a 60\% variation in six hours.  In December 1992,
\cite{1995psu..rept.....N} also reported a flare observed by IUE,
during which $S$ increased by 50\% in half an hour. In
Fig. \ref{fig.hd22468_flare} we plot the values of $S$ obtained from the
IUE spectra during these three flares. 

On the other hand, long-term time series of optical photometry were
extensively analysed for this star (\citealt{1995ApJS...97..513H},
\citealt{2006A&A...455..595L}). Recently, \cite{2007ApJ...659L.157B}
analysed photometric observations of HD 22468 in the V- band for the
years 1975-2006 and  they confirmed an activity
cycle with a period of 15-16 years,   first found in this star by
\cite{1995ApJS...97..513H}. Also from photometric observations
obtained between  1975 and 2001,
\cite{2006A&A...455..595L} found a long-term activity cycle with a
period of 19.5 $\pm$ 2.0 years.

In order to search for long-term periodic chromospheric variations, we
obtained the mean annual $\langle S\rangle$ of the indices plotted in
Fig. \ref{fig.hd22468}, excluding those points related to the
flares mentioned above. The uncertainties of each point $S$ were
obtained considering a 10\% and a 25\% error for the high and
low-resolution fluxes respectively.

\begin{figure}[htb!]
\centering
\subfigure[\label{fig.hd22468_per}]{\includegraphics[width=0.48\textwidth]{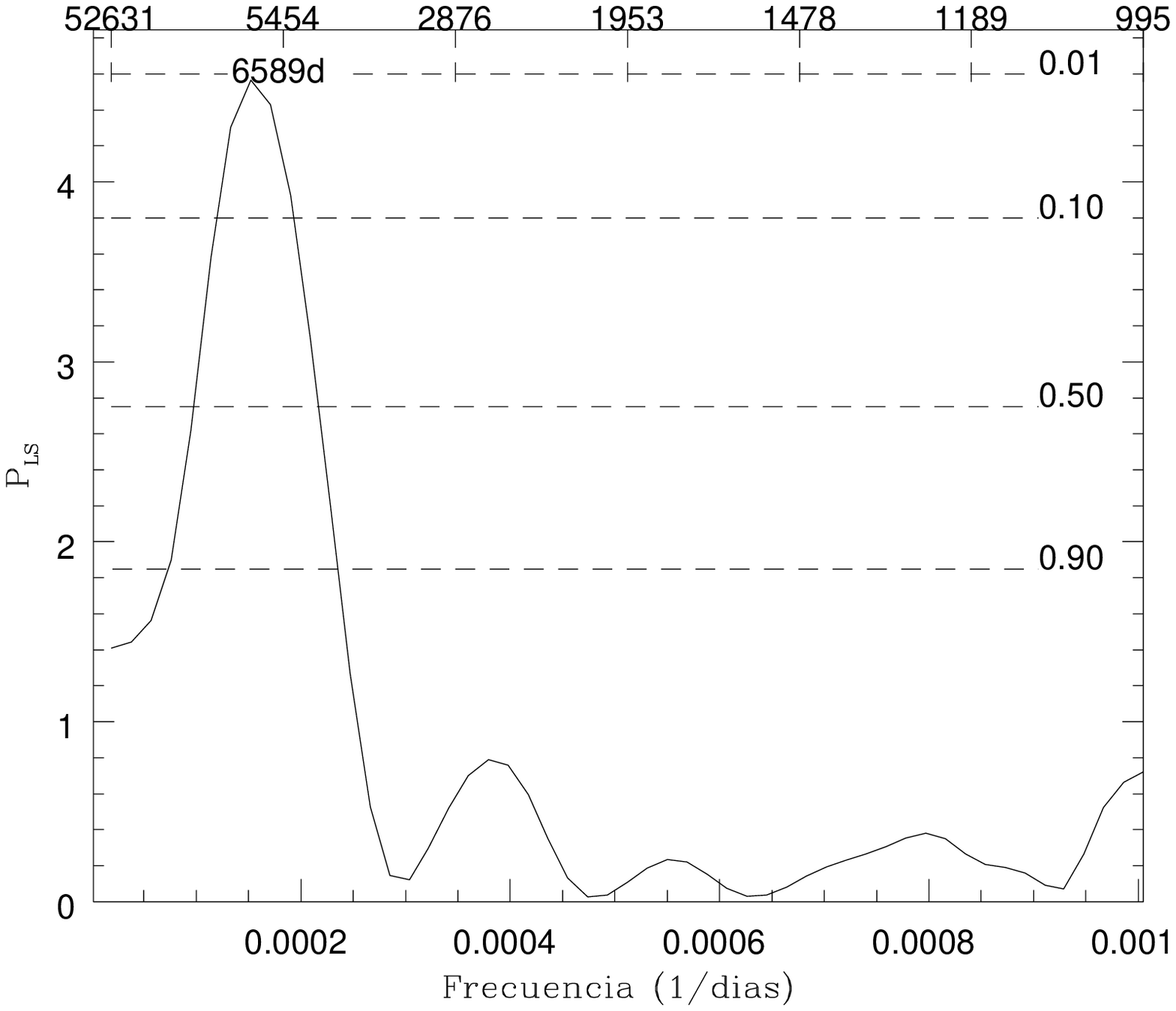}}\hfill
\subfigure[\label{fig.hd22468_fas}]{\includegraphics[width=0.5\textwidth]{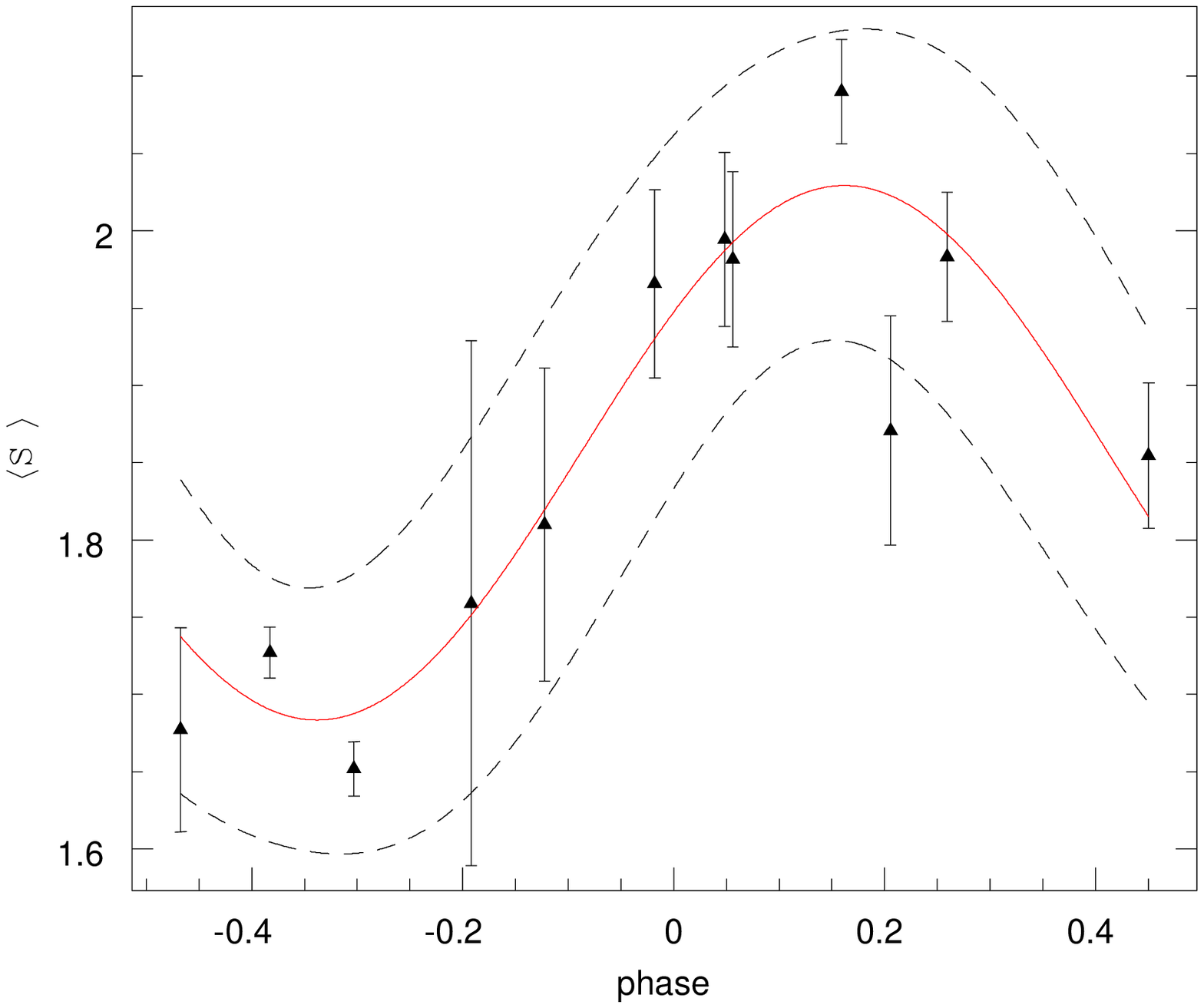}}
\caption{HD 22468 - V711 Tau - HR 1099.  \emph{Left}: Lomb-Scargle periodogram of the mean annual
$\langle S \rangle$ of the data plotted in
Fig. \ref{fig.hd22468}. The False Alarm Probability (FAP)
levels of 1, 10, 50  and 90\% are indicated. \emph{Right}: The mean annual $\langle S \rangle$
  of the data plotted in Fig. \ref{fig.hd22468} phased with the period
  of 6589 days. The solid line shows the harmonic curve that best fits the
  data with a reduced $\chi^2=1.5$ and the dashed lines indicate the points
  that apart $\sigma$ from that fit. }
\end{figure}

We analysed the indices $\langle S\rangle$
as a function of time with the Lomb-Scargle periodogram
(\citealt{1982ApJ...263..835S}, \citealt{1986ApJ...302..757H}), using
the algorithm given by \cite{1992nrfa.book.....P}. This periodogram is
shown in Fig. \ref{fig.hd22468_per}, where it can be seen that there
is, indeed, a peak at 6589 days ($\pm$1170 days) with power
$P_{LS}=4.56$ and a False Alarm Probability (FAP) of 11\%. This FAP
was computed following \cite{1986ApJ...302..757H}, who assumed a
strictly periodic signal with Gaussian noise. This condition, however, is not actually
met in this case. 

Therefore, we also computed the FAP of this period with a
Monte Carlo simulation, with an algorithm similar to the one used by
\cite{2002A&A...396..513H}. We built 10000 random series from our
dataset and we computed the Lomb-Scargle periodogram for each random
series.  Only in 110 of these periodograms a period was ``detected''
with a power $P>P_{LS}=4.56$. Therefore, in this way we obtained a
FAP=1.1\% for the peak of 6589 days in Fig. \ref{fig.hd22468_per}. In
this figure we also show the 10, 50 and 90\% FAP levels computed with
this method.

This period of $18.05\pm 3.21$ years is in concordance, within the
statistical error, with the ones found by the authors cited above.

In Fig. \ref{fig.hd22468_fas}, we plot the mean annual
$\langle S \rangle$ phased with
the period obtained  and we found that a harmonic function
fits these points with a reduced $\chi^2=1.5$, which corroborates the
period found.

\cite{2007ApJ...659L.157B} also found a 5-year flip-flop cycle,
 derived from the cyclic pattern of the peak-to-peak V
 magnitude which reflects the non-axisymmetric redistribution of the
 spotted area.  The photometric observations analysed by \cite{2006A&A...455..595L} also reflected this short-term
 oscillation with a period of 3-5 years superposed on the long-term
 spot activity cycle.

\begin{figure}[htb!]
\centering
\includegraphics[width=0.7\textwidth]{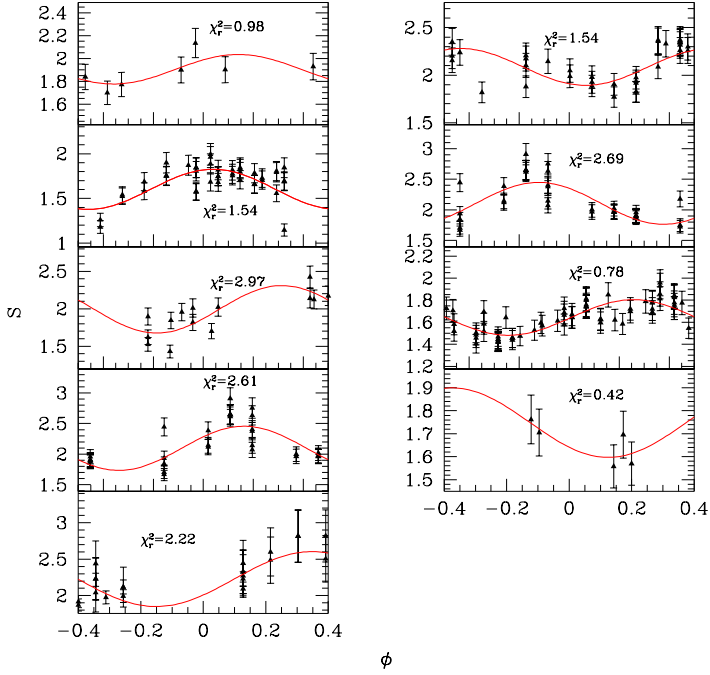}
\caption{Variability on HD 22468 - V711 Tau - HR 1099. Light curves $S$ vs. $\phi$ of the 
    datasets indicated with arrows in Fig. \ref{fig.hd22468}. We plotted the
    harmonic curve that best fit the data, the number in each plot
    refers to the reduced $\chi^2$ of the fit. }\label{fig.hd22468_curvas}
\end{figure}

To search for a chromospheric flip-flop cycle, we analysed the
peak-to-peak Mount Wilson index for several seasons of measurements,
since these variations are proportional to the difference of the area
of plages between opposite hemispheres of HD 22468.  To do this, we
built individual light curves for the sets of data indicated with
arrows in Fig. \ref{fig.hd22468}, which are shown in
Fig. \ref{fig.hd22468_curvas}. We excluded from these datasets those
points associated with flares. Following \cite{2007A&A...474..345D},
we phased each light-curve with the 2.837-day rotation period and we
fitted the data of each light-curve with a harmonic function expressed
as:

\begin{equation}
a_0+a_1 \cos(2\pi\phi)+a_2 \sin(2\pi\phi),
\label{eq.harmun}
\end{equation}
where the amplitude $A$ of the curve and its error $\sigma_A$ were computed with these
expressions:
\begin{eqnarray}
\lefteqn{A=\sqrt{a_1^2+a_2^2}\,,}\label{eq.amplitud}\\
\lefteqn{\sigma_A=A^{-1}[a_1^2\sigma_{a_1}^2+a_2^2\sigma_{a_2}^2+2a_1a_2cov(a_1,a_2)]^{1/2},}
\end{eqnarray}
where $\sigma_{a_{1,2}}$ is the standard deviations of  $a_{1,2}$ and $cov(a_1,a_2)$
is the non-diagonal term of the covariance matrix of  $a_{1}$ and $a_{2}$.

We analysed the variation of the amplitude against time, using the
Lomb-Scargle periodogram, and we obtained  that the variability of $S$
presents a cyclic behaviour with a period of 1207$\pm$ 45 days
($\sim$ 3.3 years) with a 5\% Monte Carlo FAP, which is
consistent with the period obtained by \cite{2006A&A...455..595L}.


\subsection{HD 21242 - UX Ari}\label{subsec.hd21242}
HD 21242 is a RS CVn-type system composed of a K0IV star and
a G5V companion in a 6.483-day orbit \citep{2001A&A...370..974D}. There is also a third  component, probably in a longer-period
orbit around the central binary system
\citep{1991LNP...380..297V}, which is weakly
present in the spectrum \citep{2001A&A...370..974D}.

By separating the Ca \II K line-core contributions due to each
 component of the HD 21242 system, \cite{2003A&A...402.1043A}
 concluded that the secondary star does show signatures of
 chromospheric activity, but this activity does not influence the
 average photospheric line-depth. To corroborate if this conclusion is
 also valid for chromospheric UV lines, we estimated the Mg \II
 line-core contributions of the primary and secondary stars to the
 total flux by fitting each line with two Gaussian functions, as we
 did for HD 22468 (Fig. \ref{fig.gaussfit}). In this case, however, we
 found that the contribution of the G star is larger and can be up to
 20\%. Since we could not separate the contribution of both
 stars in all spectra, we included the total flux in the analysis of
 this system.

\begin{figure}[h!]
\centering
\includegraphics[width=0.95\textwidth]{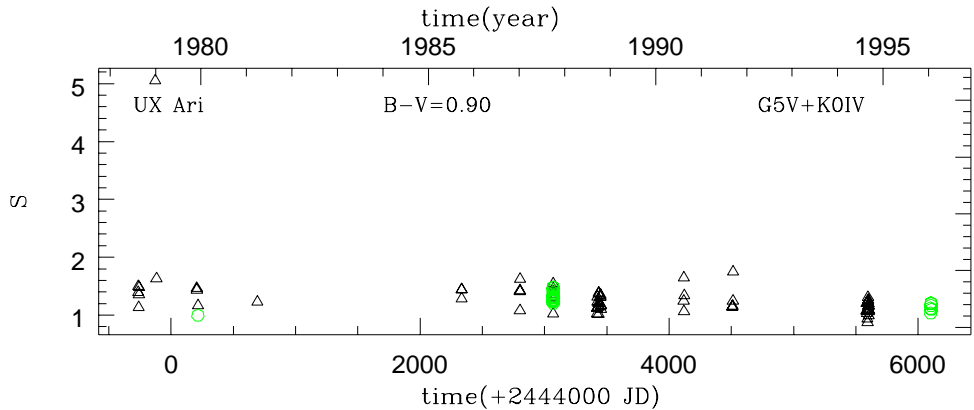}
\caption{HD 21242 - UX Ari. Symbols as in Fig. \ref{fig.hd22468}. }\label{fig.hd21242}
\end{figure}

In Fig. \ref{fig.hd21242} we plot the index $S$  obtained from all the
IUE rebinned and low-resolution spectra between 1978 and 1996 for both
primary and secondary stars. The mean annual Mount Wilson index is
$\langle S\rangle$=1.28 during this period.

 We also observe in this figure appreciable short-scale ($\sim$ months)
variations from 15\% to 80\%. In particular, during the large flare of
1979, January 1$^\textrm{st}$ \citep{1980ApJ...239..911S} the 
$S$-index reached a  value $\sim$300\% greater than the mean. From our data,
we also obtained that the index $S$ increased $\sim$60\% during the
flares of 1987, January 6$^\textrm{th}$ \citep{1988ApJ...328..610L} and
1991, September 14$^\textrm{th}$ \citep{2002ApJ...570..799S}.

\begin{figure}[htb!]
\centering
\subfigure[\label{fig.hd21242_per}]{\includegraphics[width=0.48\textwidth]{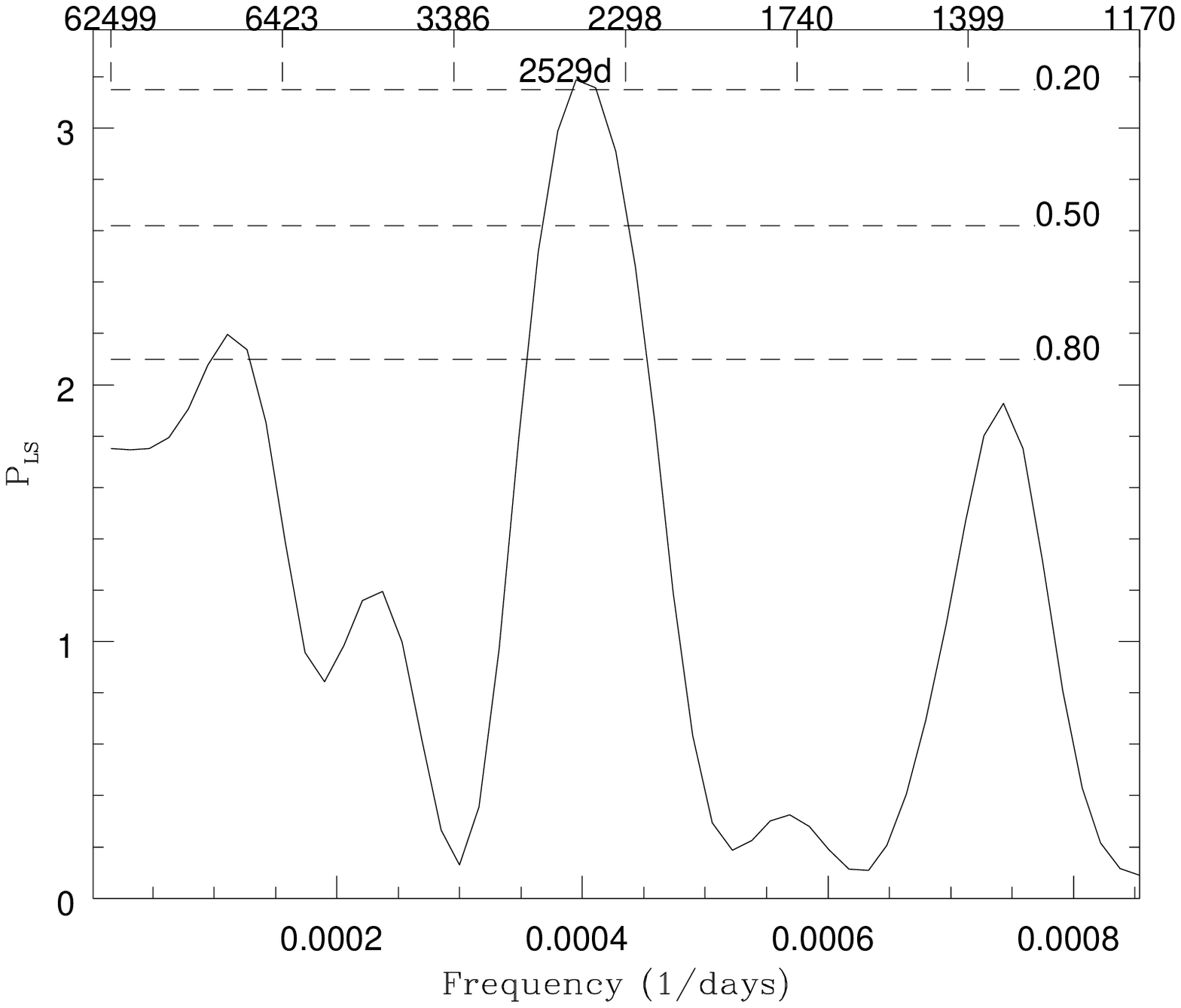}}\hfill
\subfigure[\label{fig.hd21242_fas}]{\includegraphics[width=0.5\textwidth]{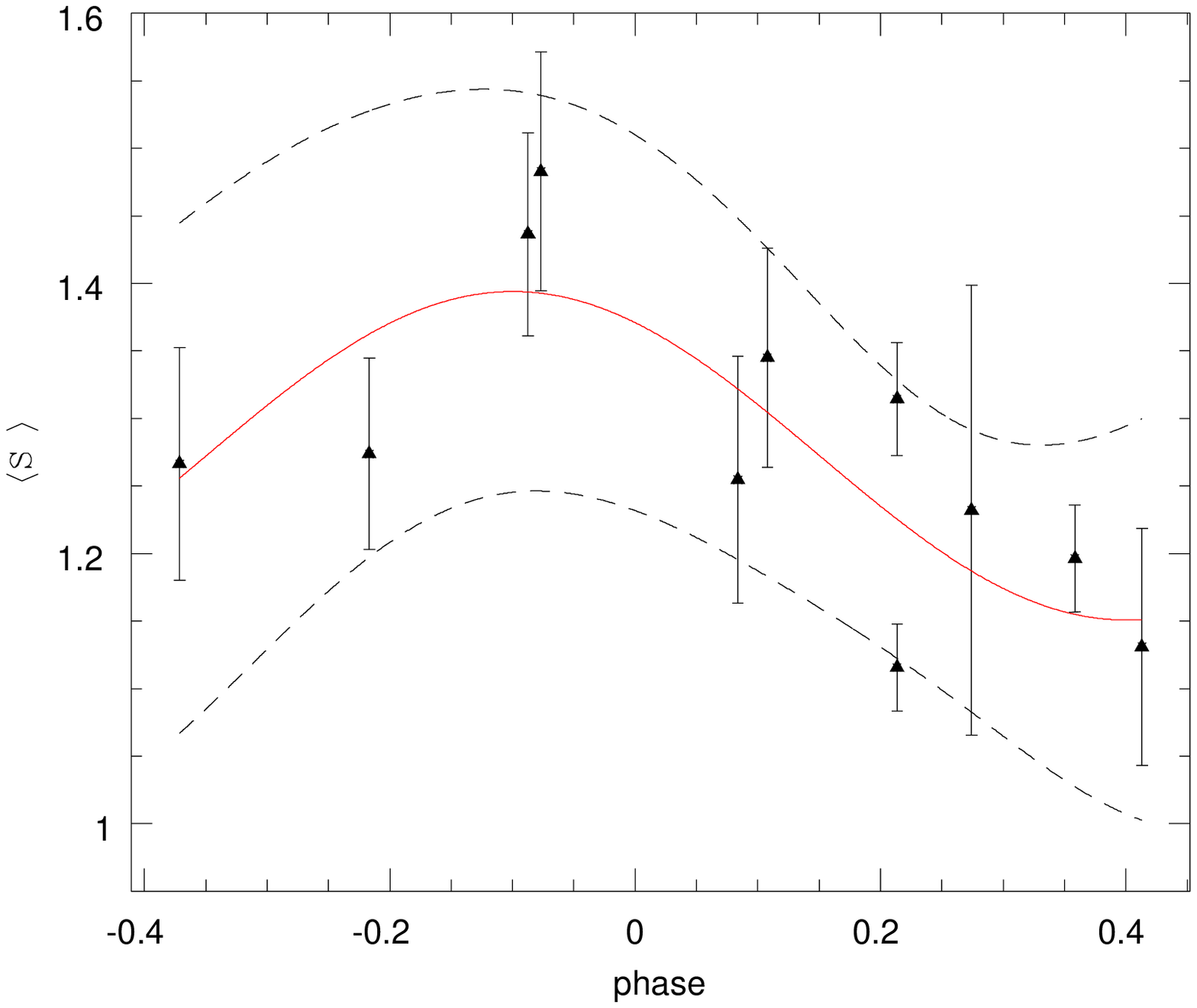}}
\caption{HD 21242- UX Ari.  \emph{Left}: Lomb-Scargle periodogram of the mean annual
$\langle S \rangle$ of the data plotted in
Fig. \ref{fig.hd21242}. The False Alarm Probability (FAP)
levels of 20, 50 and 80\% are indicated. \emph{Right}: The mean annual $\langle S \rangle$
  of the data plotted in Fig. \ref{fig.hd21242} phased with the period
  of 2529$\pm$ 101 days (6.95$\pm$0.5 years). The solid line shows the harmonic curve that best fits the
  data with a reduced $\chi^2=3$ and the dashed lines indicate the points
  that apart $\sigma$ from that fit. }
\end{figure}

In order to search for a long-term periodic chromospheric pattern, we
first obtained the mean annual $\langle S\rangle$ of the indices
plotted in Fig. \ref{fig.hd21242}, excluding those points related to
the flares mentioned above. We analysed the indices $\langle S\rangle$
as a function of time with the Lomb-Scargle periodogram, which is
shown in Fig. \ref{fig.hd21242_per}. It can be seen that there is a
peak at 2529 days ($\sim$7 years) with a  FAP of
18\%, computed with the Monte Carlo method explained in the previous section.

To our knowledge, this period has not been reported in the literature
before. However, \cite{2003A&A...402.1033A} analysed several series of
photometric observations obtained between 1987 and 2002 by different
authors and found that the mean magnitude presents a cyclic behaviour
with a period of 25 years and a significance of 99\%. Note
that a peak at 25 years is also observed in
Fig. \ref{fig.hd21242_per}, although with a FAP $>$70\%.  Conversely, in
Fig. 2 of \citeauthor{2003A&A...402.1033A}'s paper it can be found a
 peak at $\sim$5.5 years consistent with our
result with a significance close to 90\%. On the other hand, the activity cycle we obtained present a
minimum between 1980 and 1982 and a maximum between 1993 and 1995 in
agreement with \citeauthor{2003A&A...402.1033A}'s photometric
observations.

 
Given the discrepancy between both periods, we analysed
whether the IUE observations cover an interval too short and/or too
sparse to detect a cycle 25 years long. To do it, taking into account
that stellar cycles are not perfect harmonic functions, we assumed
that the star has a quasi-periodic cycle similar in shape to the solar
one, with a period of 25 years.

Following \cite{2007A&A...461.1107C}, we used the sunspot
number $S_N$ from the National Geophysical Data Center between 1900
and 2000 and rescaled the series in time to the period $P=25$ years.
\textbf{To consider data with the same signal-to-noise ratio than ours
we also rescaled the $S_N$ values and added a Gaussian noise to 
obtain a time series of the same mean
value and standard deviation of our data in
Fig. \ref{fig.hd21242_fas}.}   Then, we took a sample of the rescaled
solar data with the same phase intervals than in the IUE observations,
and we computed the Lomb-Scargle periodogram. We repeated this
procedure 1000 times with random starting dates.

\begin{figure}[htb!]
\centering
\includegraphics[width=0.6\textwidth]{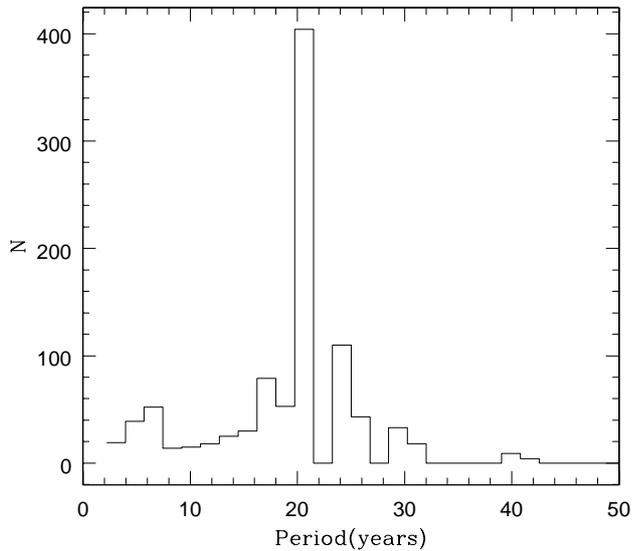}
\caption{Histogram of the detected periods (with maximum significance)
  for sample of 1000 different sets the sunspot number re-scaled to 25
  years. }\label{fig.multiperhd21242}
\end{figure}

For each periodogram, we considered the period with the maximum
significance as the detected period. In Fig. \ref{fig.multiperhd21242}
we plot the histogram of the 1000 detected periods.  In this
figure it can be seen that in 556 cases the ``correct'' period between
20 and 27 years has been detected.  Of the 444 cases where a
``false'' period is detected, 125 cases correspond to the period 18
years ($\pm$10\%), which is the time length of the dataset. In
89 cases we detected a period between 4 and 8 years. Therefore, this
test seems to suggest that our data in Fig. \ref{fig.multiperhd21242} are not
compatible with a 25-year period with a significance $>$90\%.

 Unfortunately, we are not able to detect a flip-flop cycle from the
 data plotted in Fig. \ref{fig.hd21242}, which are
 insufficient to obtain a significant period in this case.

\subsection {II Peg - HD 224085}\label{subsec.hd224085}

HD 224085 (II Peg) is a single-lined RS CVn-binary system composed by a
K2IV star and an unseen companion of an estimated spectral class M0-M3V
\citep{1998A&A...334..863B}.

\begin{figure}[h!]
\centering
\includegraphics[width=0.95\textwidth]{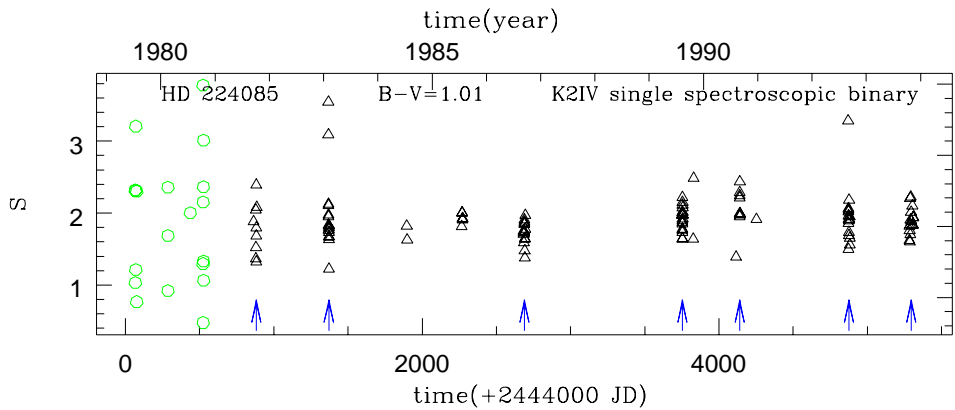}
\caption{HD 224085 - II Peg. Symbols as in Fig. \ref{fig.hd22468}.}\label{fig.hd224085}
\end{figure}

In Fig. \ref{fig.hd224085} we plot the Mount Wilson index $S$ derived
from the IUE high and low-resolution spectra obtained between 1979 and
1995. From these data, we obtained a mean level of activity given by
$\langle S\rangle$= 1.694 during this period  with
a 35\% annual variation.

 On the other hand, HD 224085 is a well-known source of flares. These
 type of events are evident in Fig. \ref{fig.hd224085} as appreciable
 short-scale variations of the index $S$. In particular, during the
 flare of 1981 October \citep{1987A&A...180..172B}, the index $S$
 increased 40\% in 18 hours. During the large flares of 1983, February
 2$^\textrm{nd}$ \citep{1988A&A...204..177A} and 1992, September 5$^\textrm{th}$ \citep{1998A&AS..127..505B}, the index $S$ reached a value 105\% and 94\%
  larger than the mean respectively.

  By analysing \textbf{25} years of photometric observations,
  \cite{2000A&A...358..624R} concluded that the spot pattern of HD
  224085 could be subdivided in two components, which present
  different cyclic behaviours. They found that one component is
  uniformly distributed in longitude and seems to be possibly
  modulated by a $\sim$13.5-year activity cycle; the other component
  is unevenly distributed and mainly concentrated in three active
  longitudes. One of these longitudes is characterized by persistent
  spot activity, but the other two active longitudes are not
  continuously active, and the spot activity switches cyclically
  back and forth between them.  The duration of the phase when only
  one longitude is predominantly active is on average $\sim$4.7
  years. This result is consistent with the 9.30-year flip-flop cycle
  previously reported by \cite{1998A&A...336L..25B} for this star.

\begin{figure}[htb!]
\centering
\subfigure[\label{fig.hd224085_per}]{\includegraphics[width=0.5\textwidth]{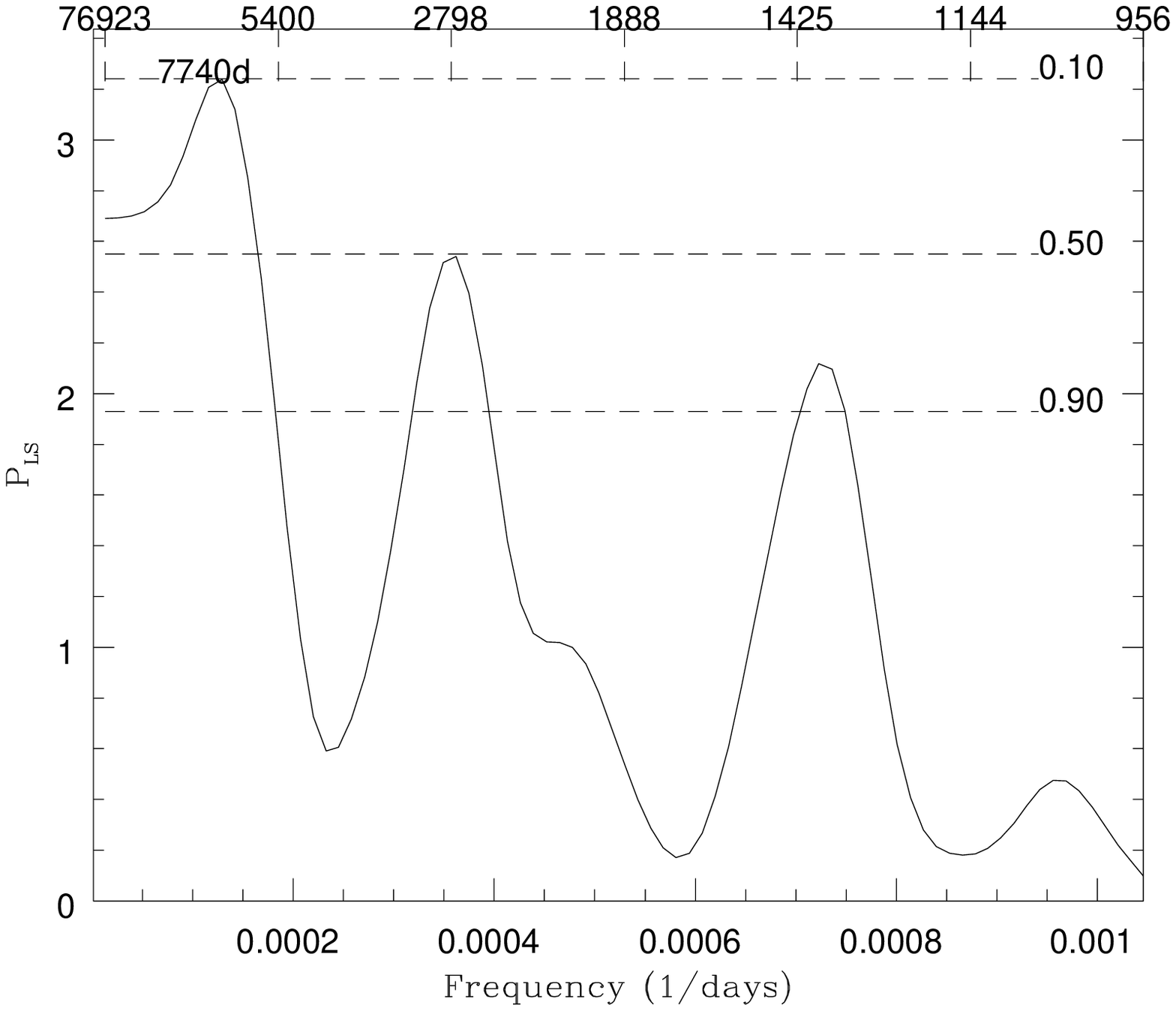}}
\hfill
\subfigure[\label{fig.hd224085_fas}]{\includegraphics[width=0.48\textwidth]{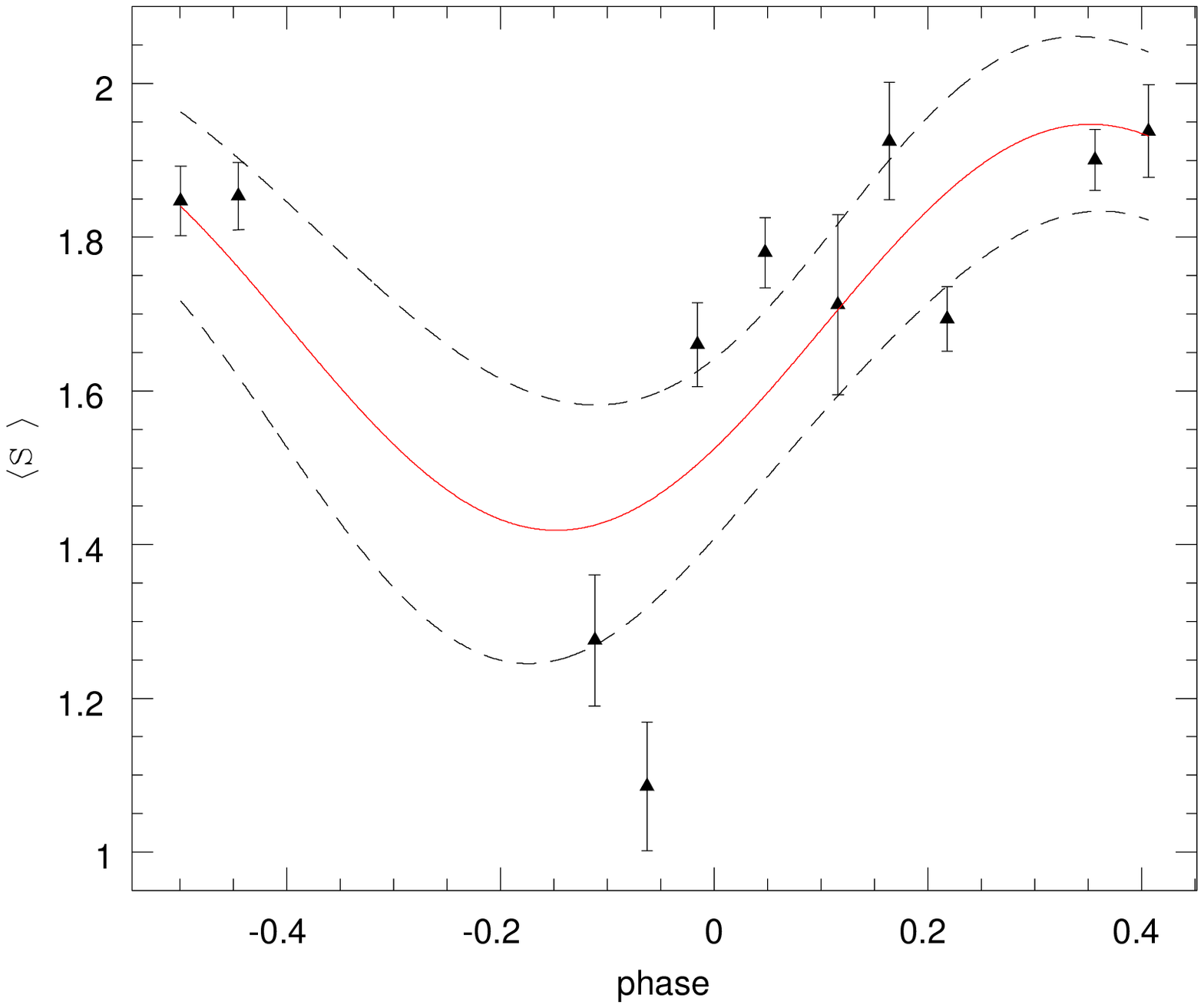}}
\caption{HD 224085 - II Peg.   \emph{Left}: Lomb-Scargle periodogram of the mean annual
$\langle S \rangle$ of the data plotted in
Fig. \ref{fig.hd224085}. The False Alarm Probability (FAP)
levels of 10, 50 and 90\% are indicated.  \emph{Right}: The mean annual $\langle S \rangle$
  of the data plotted in Fig. \ref{fig.hd224085} phased with the period
  of 7740$\pm$ 207 days (21.21 $\pm$ years 0.57 ). The solid line
  shows the harmonic curve that best fits the observations and the dashed lines indicate the points
  that apart $\sigma$ from that fit. }
\end{figure}

To search for cyclic chromospheric patterns, we first studied the mean
annual index $\langle S \rangle$ of the data plotted in
Fig. \ref{fig.hd224085} as a function of time with the Lomb-Scargle
periodogram. In Fig. \ref{fig.hd224085_per} we observe that the
periodogram has a maximum at 7740 days ($\sim$21 years) with a Monte Carlo FAP of 10\%. In Fig. \ref{fig.hd224085_fas} we show the mean annual indices
$\langle S \rangle$ phased with this period.  Although the 7740-day
peak is not sharp, only two points (with $\langle S
\rangle=1.083$ and $\langle S\rangle=1.778$)  apart more than
1$\sigma$ from the harmonic curve in
Fig. \ref{fig.hd224085_fas}. Therefore, we can only affirm that the
data of HD 224085 suggest a chromospheric cycle of near 21 years. As
far as we know, this period has not been reported in the literature
before, but an activity cycle was previously detected in this star \citep{1995ApJS...97..513H,2000A&A...358..624R}.

\begin{figure}[htb!]
\centering
\includegraphics[width=0.6\textwidth]{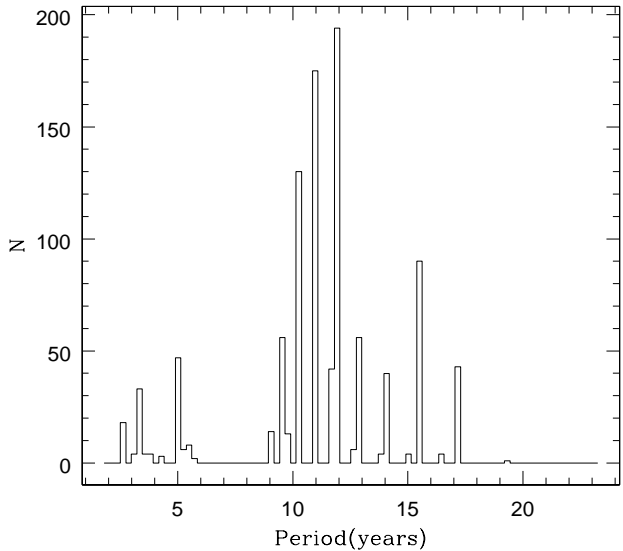}
\caption{Histogram of the detected periods for a
  sample of 1000 different sets of sunspots number re-sampled with a
  period of 12 years.}\label{fig.multiperhd224085}
\end{figure}

To check if these periods reported in the literature are
compatible with the IUE observations, we repeated the analysis done in the
previous section. We first rescaled the sunspot number to obtain a
series with the same mean value and the same standard deviation than
our data, and with a period
$P=12$ years, which is the mean period reported in the
literature. Then,
we calculated the Lomb-Scargle periodogram for 1000 subsets of
$S_N$ with the same phase intervals than our data plotted in
Fig. \ref{fig.hd224085_fas}. 

In Fig. \ref{fig.multiperhd224085} we plot the histogram of
the 1000 periods detected and we found that in 602 cases the period
detected is 12 years ($\pm$15\%).  In the histogram there is also a
peak for the period $\sim$15 years, which is the time length of the
dataset. Only in 89 of the cases a $P'\ge 16$ years ($\pm$10\%) is
detected, which is the range that include the period we obtained in
Fig. \ref{fig.hd224085_per}. Therefore, this test suggests that the data
of HD 224085 is Fig. \ref{fig.hd224085_fas} are not compatible with a
period $P=12$ years, with a confidence level $>$90\%.

\begin{figure}[htb!]
\centering
\includegraphics[width=0.7\textwidth]{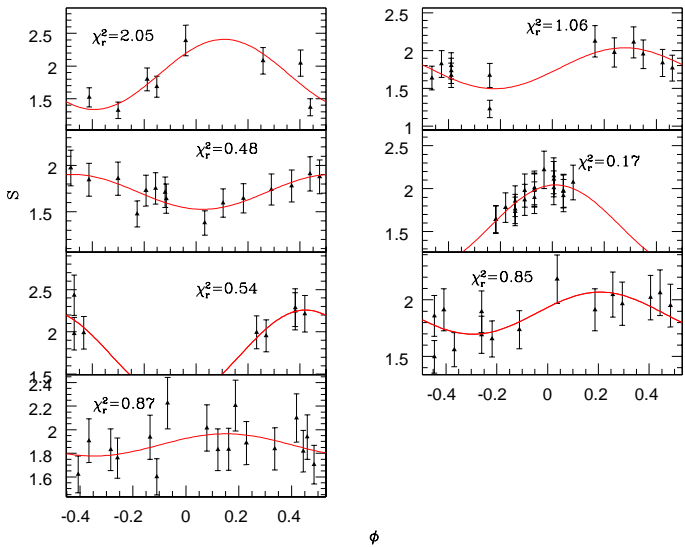}
\small{\caption{Variability on HD 224085- II Peg. Light curves $S$ vs. $\phi$ of the 
    datasets indicated with arrows in Fig. \ref{fig.hd224085}. We plotted the
    harmonic curve that best fit the data, the number in each plot
    refers to reduced $\chi^2$ of the fit. }\label{fig.hd224085_curvas}}
\end{figure}

To search for cyclic patterns of shorter periods, we finally analysed
the rotational modulation of the index $S$ during the seasons
indicated with arrows in Fig. \ref{fig.hd224085}. Similarly to the
analysis done for HD 22468, we phased each season's light-curve $S$
vs. time with the 6.724-day rotation period and we fitted each set of
data with the harmonic function given in Eq. \ref{eq.harmun}. We
analysed the amplitude $A$ (see Eq. \ref{eq.amplitud}) against time
with the Lomb-Scargle periodogram and we obtained a period of 3310
$\pm$ 253 days (9.07 $\pm$ 0.69 years) with a Monte Carlo FAP
of 7\%. This cyclic behauviour seems to be well correlated with the
flip-flop cycle of period 9.30 years obtained by
\cite{1998A&A...336L..25B} for the spot activity.

\section{Conclusions}
  
The main purpose of this paper is to study the short and long-term
chromospheric activity of the RS CVn-type stars most observed by IUE.

In the first part of this work we calibrate the Mg \II fluxes observed
in  low-resolution
spectroscopic observations available in the IUE archives with the Mount
Wilson index $S$, which allows us to use these observations for 
systematic studies of magnetic activity in late-type stars.
In particular, we found
that the Mg \II line-core fluxes derived from IUE low-resolution
spectra could be only considered as an activity indicator in stars
cooler than G3.

To complement the calibration between the Mount Wilson index $S$ and
the IUE high-resolution Mg \II line-core fluxes we obtained in Paper
I, we analysed a set of 96 nearly simultaneous observations, which
belong to 11 G3 to K3 main sequence stars, of the index $S$ and the Mg
\II fluxes derived from IUE high-resolution spectra re-sampled into
the low-resolution wavelength domain.  We obtained a $B-V$ colour
dependent relation between the index $S$ and the IUE low-resolution Mg
\II line-core fluxes.

In the second part of this work, we studied the long and short-term
chromospheric activity of the non-eclipsing RS CVn stars most observed
by IUE: HD 22468 (V711 Tau, HR 1099, K1IV+G5V), HD 21242 (UX Ari,
K0IV+G5V) and HD 224085 (II Peg, K2IV).  To do so, we analysed with
the Lomb-Scargle periodogram the mean annual index $\langle S\rangle$
and the amplitude of the rotational modulation of $S$ of each star.

We obtained a possible chromospheric activity cycle in HD 22468 with a
 period of $\sim$18 years and a chromospheric flip-flop cycle of period
 $\sim$3 years. We also found a flip-flop like cycle with a period of $\sim$9
 years in the HD 224085 data.  These results are in agreement with the 
 ones reported in the literature.  On the other hand, we also detected a cyclic
 chromospheric pattern in the mean annual index of HD 224085 and HD
 21242 with periods of $\sim$21 and $\sim$7 years
 respectively. 

\begin{acknowledgements}

We thank the referee Dr. Sergio Messina, for his useful comments and
observations.
\end{acknowledgements}

\bibliographystyle{aa}


\end{document}